# NUMERICAL INVESTIGATION OF SOOT FORMATION IN TURBULENT

# DIFFUSION FLAME WITH STRONG TURBULENCE CHEMISTRY INTERACTION

**B. Manedhar Reddy**
Department of Aerospace Engineering,
Indian Institute of Technology Kanpur,
Kanpur, Uttar Pradesh, India.
E-mail: manedhar@iitk.ac.in

**Ashoke De[1]**
Department of Aerospace Engineering,
Indian Institute of Technology Kanpur,
Kanpur, Uttar Pradesh, India.
E-mail: ashoke@iitk.ac.in
Tel.: +91-512-6797863; Fax: +91-512-6797561
ASME Membership: 000100238660

**Rakesh Yadav**
ANSYS Fluent India Pvt. Ltd.,
Pune, Maharashtra, India.
E-mail: rakesh.yadav@ansys.com

## ABSTRACT

*The present work is aimed at examining the ability of different models in predicting soot formation in 'Delft flame III', which is a non-premixed pilot stabilized natural gas flame. The turbulence-chemistry interactions are modeled using a laminar steady flamelet model (SLFM). One step and two step models are used to describe the formation, growth, and oxidation of soot particles. One Step is an empirical model which solves the soot mass fraction equation. The two-step models are semi empirical models, where the soot formation is modelled by solving the governing transport equations for the soot mass fraction and normalized radical nuclei concentration. The effect of radiative heat transfer due to gas and soot particulates is included using P1 approximation. The absorption coefficient of the mixture is modeled using*

---
[1] Corresponding author, Ashoke De

*the weighted sum of gray gases model (WSGGM). The turbulence-chemistry interaction effects on soot formation are studied using a single variable probability density function (PDF) in terms of a normalized temperature or mixture fraction. The results shown in this work clearly elucidate the effect of radiation and turbulence-chemistry interaction on soot formation. The soot volume fraction decreases with the introduction of radiation interactions, which is consistence with the theoretical predictions. It has also been observed in the current work that the soot volume fraction is sensitive to the variable used in the PDF to incorporate the turbulence interactions.*

## 1. INTRODUCTION

Combustion is a complex phenomenon and generally involves complex chemical reactions along with heat and mass transfer. Pollutant emissions from combustion like NOx and particulate carbonaceous matter (soot) are a major concern. Pollutants are not the unburned hydrocarbon, they are the hydrocarbon produced during combustion and not consumed by the flame. Soot is the condensed carbon particles resulting from the incomplete combustion of the hydrocarbons, which indicates poor utilization of fuel. One of the major concerns with theoretical prediction of soot is the complex physics of soot formation, which is poorly understood and makes modeling a difficult task. Intensive theoretical and experimental investigations in the last two decades have helped in improving the understanding of soot and have led to a detailed insight into the evolution of soot formation process. Haynes and Wagner [1], Kennedy [2] and Bockhorn [3] reviewed these efforts. Nucleation is the starting step of soot formation, which takes place due to condensation of polycyclic aromatic hydrocarbon (PAH) species and results in formation of small particles. The growth of these nucleated small particles depends on the heterogeneous surface reactions with the gas phase, primarily acetylene, which is one of the main species involved in the surface growth. One of the most widely used approach to model the surface growth is the H-Abstraction-Carbon-Addition (HACA) mechanism [4]. The soot particles further grow by coagulation, which is due to collisions of soot particles with each other. Heterogeneous reactions with molecular oxygen and OH radicals result in the oxidation of soot.   The PAH, the building blocks of soot, exhibit a strong sensitivity to the local scalar dissipation rate in the flow field. These interactions of soot with the turbulence, molecular transport and gas-phase chemistry result in very high temporal and spatial intermittency. In addition to the type of fuel, soot formation also depends on the operating and prevailing flow conditions. Therefore, the soot formation can be reduced by controlling operating parameters like residence time, temperature and the turbulence. If the time to react at high temperature is more, it will result in oxidation of soot and hydrocarbon, but at the expense of an increase in NOx formation.

The modeling of soot formation in turbulent flames has spanned from RANS to LES to DNS. Most of the studies have employed empirical or semi-empirical soot models and neglected the soot-turbulence-chemistry interactions. The detailed modeling of soot formation in turbulent flames is an extremely challenging task. The resolution of the chemistry of soot formation typically involves a large number of reactions involving stiff chemistry. Further, the modeling of the turbulence chemistry interaction requires the resolution of an evaluation of a high dimensional PDF [5] and hence expensive. In order to solve the gas phase reactions, the mixture fraction based methods [6] are commonly employed. Once the acetylene is formed, the process of polymerization is typically very fast, therefore the modeling of soot formation can be significantly simplified by considering acetylene as the primary soot precursor. Modeling the effect of radiation on soot formation is complex as it affects the soot kinetic rate associated with precursors of soot. Modeling of soot-turbulent-chemistry interactions is a highly demanding task as soot is confined to very thin structures due to a very large Schmidt number. Recently Bisetti et al. [7] carried out a DNS study on the evolution of soot and Michael et al. [8] used LES to investigate the soot-turbulence-chemistry interactions effect on the soot formation process of Delft flame III. In the previous studies [2], soot-radiation interactions were not considered and also the temperature based PDF had not been used to incorporate the soot-turbulence-chemistry interactions. The primary objective of this study is to explore semi-empirical soot models to predict the soot formation in a turbulent flame along with strong turbulence chemistry and turbulence radiation interactions. The computational results along with experimental studies can be used to enhance the knowledge of the soot evolution process. In the current work, a kinetic based model along with the flamelet model is used to describe the formation and oxidation of soot. The soot particulate effect on absorption coefficient of the medium is considered to incorporate the coupling between soot and temperature. The soot-turbulence interactions are incorporated using a presumed shape PDF. The effect of the soot-radiation interactions and the small scale interactions between particle dynamics, chemistry and turbulence on the formation, growth and oxidation of soot are examined. Different models are used for the prediction of soot particles and the results are discussed and compared with the experimental data of Qamar et al [9].

## 2. THEORETICAL DESCRPITION OF LAMINAR STEADY FLAMELET MODEL

The flame surface is defined as an iso-surface of the mixture fraction for non-premixed combustion [10-15]. The conserved scalar model equations for equilibrium chemistry are used here, but with a chemical source term. The flame equations are transformed from physical space to mixture fraction space ( $f$ being the independent variable) where N equations are solved for the species mass fraction and one equation for the temperature in the mixture fraction space.

$$\frac{\partial Y_i}{\partial t} = \frac{1}{2}\chi\rho\frac{\partial^2 Y_i}{\partial f^2} + S_i \tag{1}$$

$$\rho\frac{\partial T}{\partial t} = \frac{1}{2}\rho\chi\frac{\partial^2 T}{\partial f^2} - \frac{1}{C_p}\sum_i H_i S_i + \frac{1}{2C_p}\rho\chi\left[\frac{\partial C_p}{\partial f} + \sum_i C_{p,i}\frac{\partial Y_i}{\partial f}\right]\frac{\partial T}{\partial f} \tag{2}$$

The scalar dissipation rate is modelled across the flame and is given by

$$\chi_{st} = \frac{a_s exp(-2[erfc^{-1}(2f_{st})]^2)}{\pi} \tag{3}$$

For adiabatic systems, the temperature and species mass fraction of the laminar flamelets are parameterized by $f$ and $\chi_{st}$. The PDF of $f$ and $\chi_{st}$ is used to determine the temperature and mean species mass fraction in turbulent flames as:

$$\bar{\phi} = \int\int\phi(f,\chi_{st})p(f,\chi_{st})dfd\chi_{st} \tag{4}$$

## 2.1 Radiation Modeling

The WSGG model is implemented by considering total emissivity over a distance $s$ as Eqn (5). Four fictions gases are considered to represent the non-gray emission of the mixture. The details of the approach can be found in Rakesh et al. [16].

$$\epsilon = \sum_{i=0}^{I} a_{\epsilon,i}(T)(1-e^{\kappa_i p s})$$

(5)The radiative transfer equation is simplified based on the optically thick medium approximation into a truncated series expansion in spherical harmonics (P1 approximation) [17] given by Eqn (6)

$$-\nabla \cdot (\frac{1}{3(K_{av}-K_{sv})-Ak_{sv}}\nabla G_v) = K_{av}(K_{bv}-G_v)$$

(6)

## 2.2 Soot Modeling

In the one-step model [18], a single transport equation is solved for the soot mass fraction Eqn (7)

$$\frac{\partial \rho Y_{soot}}{\partial t} + \nabla \cdot (\rho \overline{v})Y_{soot} = \nabla \cdot (\frac{\mu_t}{\sigma_{soot}}\nabla Y_{soot}) + R_{soot}$$

(7)

Where $R_{soot}$ the net rate of soot generation is calculated using the balance of soot formation and combustion. In the Two- Step model [19], in addition to the above Eqn 7, the normalized radical nuclei concentration is also solved.

$$\frac{\partial \rho b_{nuc}^*}{\partial t} + \nabla \cdot (\rho \overline{v})b_{nuc}^* = \nabla \cdot (\frac{\mu_t}{\sigma_{nuc}}\nabla b_{nuc}^*) + R_{nuc}^*$$

(8)

In the Moss-Brookes model [20] the soot mass concentration is calculated instead of $R_{soot}$ in Eqn 7 and the instantaneous production rate of soot particles is calculated instead of $R_{nuc}^*$ in Eqn 8. For the One Step and Moss-Brookes models, only methane as fuel was considered to determine the model constants. For the Two-step model, propane as fuel was used to determine the pre-exponential constant as it is applicable to other hydro carbons also.

*2.2.1 Soot radiation* interaction: Presence of soot particulates can greatly affect radiative properties like absorption coefficient of the medium. The effective absorption coefficient of soot is used by the soot models to determine the effect of soot on radiative heat transfer. In the current work, sum of the absorption coefficient of gas and soot is used as the effective absorption coefficient of the medium.

$$\kappa_i = k_i + k_{soot} \tag{9}$$

The absorption coefficient due to soot particulates is calculated as a function of soot concentration [2].

*2.2.2 Turbulence chemistry interaction:* Moss-Brookes model calculates the soot concentration by computing a time-averaged soot formation rate from the averaged flow-field information at each point in the domain [22]. Theoretical description of the turbulent flow can be accurately predicted by PDF method. In the Moss-Brookes model, a single-variable PDF normalized in terms of the mixture fraction or temperature is used to incorporate the effect of turbulence on soot formation. Transport equations are solved to obtain the mean values of the independent variables required for the construction of the PDF. In the current work, the turbulence chemistry interactions for calculating the mean soot and nuclei source terms are considered only for the two equation Moss-Brookes soot model.

## 3. NUMERICAL DETAILS

In the current work, all computations are performed using ANSYS FLUENT 13.0 [23]. For all equations, a second order discretization scheme is used. SIMPLE algorithm is used for pressure-velocity coupling. In order to model the turbulent flame, three different turbulence models, Modified k-ε model (MSKE) [24], Realizable k-ε model (RKE) [25] and Reynolds stress model (RSM) [26] are used. The turbulence chemistry interactions for the gas phase chemistry are modeled using SLFM [10] model. The chemistry is represented by GRI 3.0 chemical mechanism [27].

## 4. DESCRIPTION OF TEST CASE

The test case used here is Delft Flame III, it is one of the flame used in Turbulent Non-premixed Flame (TNF) workshops [28]. Peeters et al. [29] have provided the details of the burner. The diameter of the fuel jet is 6 mm and a rim with an external diameter of 15 mm surrounds it. The rim acts as a bluff body which generates a recirculation of hot gases and stabilizes the flame. The rim has 12 equally spaced holes at a radius of 3.5 mm and diameter of 0.5 mm. The holes act as pilot fuel inlet, hot equilibrated mixture enters the domain with a power of ~200 W which accounts for ~1% of the fuel jet power. There is a concentric circular jet of external diameter 45 mm for the primary air. The outer most concentric face acts as secondary air inlet. The experimental test conditions are listed in Table 1.

The most challenging part of modeling the flame is the pilot flames which control the flame stabilization at the exit of the burner. Naud et al. [30] suggested an approach in which the pilot flame inlet is modified by increasing the mass flow rate of the pilot. This approach is not strictly required in SLFM modeling. However, in order to have consistency with previously published work, we have also used the same approach for pilot flow in the current work.

## 5. MODELING DETAILS AND BOUNDARY CONDITIONS

Because of flow symmetry, an axisymmetric 2D grid has been used for the current computational work. The equivalent area of the 12 holes is used to calculate the diameter of the pilot fuel inlet. The computational domain extends by 250D and 50D, where D is the diameter of the fuel jet (Fig 2). The fuel is natural gas and is assumed to consist of $CH_4$, $C_2H_6$ and $N_2$ only. The mass fractions are adjusted such that it has the same calorific value as that of the original fuel. Therefore, the fuel contains 81% $CH_4$, 4% $C_2H_6$ and 15% $N_2$ by volume.

The axial and radial velocity profiles were modified to avoid spreading in the upstream direction. The velocity conditions at X = 3mm were used, the magnitude of the profiles were modified while maintaining the shape to achieve the experimental mass flow rate. This approach has been used in earlier [29-31] with this burner. The turbulent kinetic energy, turbulent dissipation and Reynolds stress components for the flow inlet are taken from experimental data [28].

## 6. RESULTS AND DISCUSSION

The flamelet computations were performed and presented in this section. The ability of different turbulence models to capture the jet spreading is studied. The effect of radiative heat transfer on the flame temperature is also investigated, followed by a qualitative study of the soot formation. Soot radiation and soot turbulence-chemistry interactions have been extensively studied and reported in details.

## 6.1 Grid Independence

Two non-uniform grids of 200×150 and 400×300 points are used to investigate the grid independence for the current work. The MSKE turbulence model is used for the simulations.  The boundary conditions as described in the Table 1 are used. Fig 3 shows the predictions using the two different grids and the comparison with the experimental results. The results obtained using both the grids are in good agreement with each other; thus the coarse grid (200×150) has been chosen for the rest of the simulations reported in the current computational work where the fuel inlet, pilot fuel, pilot wall rim are resolved using 13, 12 and 6 grid points, respectively in the coarse  grid.

## 6.2 Velocity

Based on the work of Naud et al. [30], mixture fraction is used to represent the pilot flame with a value of 0.115. The mean pilot velocity of 8.28 m/sec has been used as also been used earlier [30, 31]. The velocity near the jet has a significant effect on the flame stability. Bearing this into account, three turbulence models, namely MSKE, RKE and RSM have been used to compare the predictions. In MSKE and RSM, $C_{\varepsilon 1}$  has been adjusted to 1.6. The radial profiles of velocity and turbulent kinetic energy at different axial locations and using different turbulence models are shown in Fig. 4. At the upstream location (X=50 mm), the predictions by all the models are in excellent agreement, but as we move downstream RSM shows better predictions compared to the other two. RKE model is completely unable to capture the proper spreading of the jet, primarily due to the improper balance of production and dissipation terms in the turbulence models. This under prediction of velocity can be associated with the over prediction of turbulent kinetic energy as can be inferred from Fig. 4. The over prediction of turbulent kinetic energy results in large values of turbulent shear stresses and eddy viscosity, so that the central jet decelerates too quickly and spreads too much. Both the velocity and the turbulent kinetic energy predictions by RKE model appear to be the worst amongst all, whereas the velocity predictions by MSKE and RSM models are quite comparable with each other and are in good agreement with experimental values. It is observed that the turbulent kinetic energy over predictions near the burner is much lower with the MSKE. The turbulent kinetic energy is found to be in good agreement with the experimental data as we go downstream, and this can be associated with the under prediction of mean velocity. However, the spreading rate and the turbulence intensity remain over-predicted by MSKE. The RSM predictions show an improvement along the radial direction. The predictions of turbulent kinetic energy tend to improve along the downstream. The slight under-estimation of velocity at the downstream positions is due to the over-estimation of mean temperature; the mean axial velocity is low due to increase in diffusion. The RSM predictions of the current model differ from the results obtained by Rakesh et al. [31], the main difference is that we have used steady flamelet while they have used joint composition PDF based model as the hand-shaking between turbulence model and chemistry is done through the density, which in turn affects the mean velocity and kinetic energy.

**6.3 Temperature**

Radial profiles of the temperature are shown in Fig. 4. RKE predictions are consistently higher near the axis. The predictions of MSKE and RSM are relatively more accurate and have closer matches with the experimental values, while RKE model is unable to capture the temperature profiles at the downstream locations. With RSM and MSKE, although the shape of the temperature profile is in good agreement near the axis but slightly higher predictions are observed for the radial peak temperature. Despite the higher prediction of mean mixture fraction by RSM and MSKE which should result in lower mean temperature, the temperature is over predicted. A similar comment can be made about the agreement of the temperature that is observed above the axis despite over prediction of the mean mixture fraction.The inconsistencies can be attributed to the combustion model and can be explained by the fact that the temperature is over-estimated by the flamelet model. This also explains the over prediction of temperature near the axis by RKE. The profiles of Fig. 4 indicate that the flame length is slightly over-estimated by all the turbulence models in order of RKE > MSKE > RSM. The experimental data is available for the region close to the burner head; X=300 mm is maximum downstream distance where data is available Therefore no compelling statement can be made about the accuracy in the predictions of the flame length by different turbulence closures.

The effect of radiative heat transfer for this flame has been investigated by Rakesh et al. [31] by using joint composition PDF and Habibi et al. [33] using flamelet modeling. Both observed that the temperatures reduced by ~100K at upstream locations as ~ 250 mm with inclusion of radiative heat transfer. In the current work, the effect of radiation has also been studied using the P1 model in conjunction with RSM for turbulence modeling. The temperature contour with different radiation modeling approaches is shown in Fig. 5. A similar trend is observed here, where the peak centerline temperature dropped by ~ 100K and shifted to upstream by ~ 3D with gray radiation; while with the non-gray radiation, the peak centerline temperature is dropped by another ~350K and shifted by ~6D in the upstream direction with non-gray radiation. The global maximum temperature has also dropped from 2090K to 2040K with gray radiation and dropped to 1980K with non-gray radiation. The location of global maximum temperature has shifted away from the centerline in the radial direction by ~2D with gray radiation and ~4D by non-gray radiation. The radial location of global maximum temperature has not been effected by radiation.

**6.4 Mixture fraction**

Radial profiles of the mean mixture fraction are depicted in Fig. 4. The convection–diffusion equation governs the evolution of the mean mixture fraction, so it can be linked with the prediction of spreading rate by different models [29]. While analyzing the experimental uncertainty, Nooren et al. [32] reported that there is typically 5% uncertainty for mixture fraction measurements and a maximum of 20%. The RKE predictions are attractive near the axis, but exhibits radial broadening at the downstream locations. This is due to over prediction of eddy viscosity which results in excess diffusion about the axis and counter acts the excess mean density prediction. The large amount of diffusion compensates the over prediction on the axis. It can be noted that the MSKE model over predicts the mean mixture fraction value near the axis due to the under prediction of eddy viscosity near the axis, whereas the RSM model also over predicts the mixture fraction, as observed at X = 100mm and X= 250 mm. The over-estimation can be associated with the over estimation of mean density. This results in an excessive convection at the axis, which translates into over prediction of the mean mixture fraction. Moreover, the mixture fraction in the inner shear layer (r<10 mm) is over-estimated by both the MSKE and RSM models, which is primarily due to the improper balance between convection and diffusion in 'mean mixture-fraction' equation.

## 6.5 Species Mass Fraction

The radial profiles of the major products of combustion are reported in Fig. 6. The discrepancies observed with other quantities are consistent with the discrepancies observed for all the species. For example, the estimation of CO and $H_2$ about the centre line is over-predicted where the mixture fraction is also over predicted. The predictions of CO are reasonable and consistent with the temperature predictions.  However, OH is over-estimated, while the major species $H_2O$ is very well reproduced. Agreement with the experimental result is good as detailed mechanism (GRI 3.0) has been used for representing chemistry. Taking into account the large uncertainties in the experimental predictions for individual species mass fractions, the predictions are in good agreement with the experimental measurements.

## 6.6 Soot Predictions

Prediction of soot with different modeling approaches is one of the prime motivations for the current study. The RSM turbulence model is used for soot predictions as it gives good reproduction of the spreading rate. Here, the soot computations are done using three different soot modeling approaches. Fig. 7 shows the comparison of the axial profiles of the centerline soot volume fraction using the different soot models with the experimental measurements of Qamar et al [9] and LES predictions of Michael et al. [8]. The one-step soot model under predicts the soot volume fractions at lower axial locations and is followed by a steep peak towards the downstream. In case of one step model, the soot formation rates are proportional to temperature. Since the temperature is over predicted in the current predictions at higher downstream locations, the soot formation also shows a peak there.    The peak is followed by rapid oxidation as it passes through the iso-surface of the stoichiometric mixture fraction. The oxidation rates in one-step model are dependent on the mass fraction of oxidizer, the turbulence levels and the mass fraction of soot itself. Due to this, the oxidation rates are over predicted as we move to the downstream. The two-step and the Moss-Brookes predictions are in better agreement with experimental values. In case of two-step model, the case without radiative heat transfer shows an excellent match with the experimental profiles, but including radiation shows under prediction. The soot formation and the nuclei generation rates in the two-step model is significantly affected by the empirical constants. So, this model may require tuning of these empirical constants.  Although the model has shown under prediction with the correct temperature field, but still it is able to predict the trend and profile correctly and hence better than the one-step model. Though the Moss-Brookes model too rely on the model constant, however the soot formation and destruction rates in this takes care of the chemistry through use of the soot precursors and oxidizers. In case of Moss-Brookes model, the soot is over predicted for the case without radiation modeling, but the predictions keeps on improving as we include the radiative heat transfer and with non-gray radiation model, the predictions matches nicely with the experimental values. In the current work, the inception and surface growth is modeled using acetylene concentration only, which tends to over predict the inception and surface growth rates. This has also been highlighted by Michael et al. [8], where they have investigated the soot formation using LES and observed considerable over predictions resulted due to significant uncertainty in the mechanism of PAH growth and formation in methane flames. In addition to the formation, the oxidation is another important aspect for the Moss-Brookes model. The results shown in Figure 7 (c) are with Lee [34] oxidation model. In order to see the impact of different oxidation models, we revisited the Moss-Brookes model with oxidation model of Fenimore-Jones [35]. Fig. 8 shows the soot predictions of the Moss-Brookes model using these two different oxidation models. The discrepancies between the two models are because of the principal oxidizer. Fenimore-Jones model uses OH radical and Lee uses both OH and $O_2$ radicals. The collision efficiency is 0.04 and 0.13 for Fenimore-Jones and Lee model respectively. Thus the oxidation is not properly predicted resulting in over prediction of soot by Fenimore-Jones model and expounds the effect of collision efficiency and oxidation species on soot formation. The contour of the soot volume fraction is shown in Fig. 9. The one step is not able to capture the soot formation in the reaction zone. As explained earlier, soot formation is predicted to occur in the primary reaction zone, this can be observed in the counter of two step and Moss-Brookes model. The contours show fewer discrepancies than the experimental data. Nevertheless, there is good agreement between experiments and predictions in terms of the soot volume fraction.

*6.6.1 Soot radiation interaction:* It is well known that the coupling between soot and radiation is strong. Soot is very much dependent on the temperature so it important to account for the heat loss in an accurate way. Radiative transfer of thermal energy from the flame is substantial increased by soot which has high emissivity. This has lowered the flame temperature. The center line maximum temperature location is not affected by soot radiation interaction. The effect of radiation on soot volume fraction can be observed from Fig. 7. All the models have shown a decrease in soot prediction with the inclusion of soot-radiative interaction. The decrease in the soot yield signifies a higher radiative loss from the flame and is consistent with the theory. Decrease in temperature affects the reaction rate of precursor formation. This result in the decrease of maximum soot volume fraction is due to the decrease of amount of precursor produced to form soot. The soot volume fraction decreases by an order of two with gray radiation and an order of six with non-gray radiation with the one step model indicating the sensitivity of soot model to temperature. The decrease of temperature in the radial direction results in decrease of soot inception which translates into decrease of soot volume fraction. The decrease in soot volume fraction prediction by Two Step model is of order of two with gray radiation and decreases by an order of ten with non-gray radiation.

In Two step model, the soot formation and oxidation is governed by the formation of nuclei. Due to decrease in temperature by the radiative transfer, not enough nuclei are formed and there is deceleration in the rate with which these nuclei are consumed to produce soot particles resulting in the decrease of the soot volume fraction. Another reason for the drastic reduction might be due the model parameters used in the models. As mentioned earlier, a parametric study is required to find the optimum parameters for the current fuel and operating conditions to improve the current predictions for two-step model.  In case of one-step and two-step models, the change in temperature directly affects the rate of soot formation or nuclei generation. However, in case of Moss-Brookes model, the effect is more complicated as the soot rates are dependent not only temperature but the species concentration of precursor and oxidizer species, which in turn also gets affected by change in temperature. Therefore, the sensitivity of Moss Brookes model with temperature is considerably low compared to the other models. The contours of soot nucleation, soot surface growth and soot oxidation using Moss Brookes model are shown in Fig. 10. It can be observed that the radiation has significant effect on the nucleation which has reduced drastically about the center line. Gray radiation results in slight reduction and with non-gray radiation the soot volume fraction decreases by a factor of two. The location of maximum soot volume fraction has shifted from ～100D to ～112D with non gray radiation due to shift in surface growth and oxidation location as can be observed from

Fig. 10. The prediction with non-gray radiation is in good agreement with experimental values.

*6.6.2 Soot turbulence interaction:* The soot-turbulence interactions were incorporated in the Moss-Brookes model and effect of turbulence-chemistry interactions on soot are shown in Figs. 11, 12 and 13. It can be observed that the soot volume fraction decreases when temperature based PDF is used and increases when the mixture fraction based PDF is used. The temperature based PDF uses over predicted mean temperature values as the limits of integration to obtain the average soot production rate. Due to pyrolysis and oxidation of soot precursors at high temperatures the soot volume fraction decreases. With the non-gray radiation also the soot volume fraction decreases with temperature based PDF due under prediction of instantaneous rate of soot production. The increase in soot volume fraction when mixture fraction based PDF is used might be due to stretching of PAH caused by mixture fraction fluctuations, leading to a higher average soot productions as observed in Fig. 11. Due to high temporal and spatial intermittency, the PDFs are not able to predict the average soot production rate accurately. The axial profiles of soot surface growth rate and oxidation rate are shown in Fig. 12, it can be inferred that both the surface growth and oxidation exhibit a similar trend as explained above.

The amount of surface growth is reduced with inclusion of temperature based PDF. The increase in surface growth rate with mixture fraction based PDF is due to increase in the instantaneous soot volume fraction prediction and the oxidation rate increases due to decrease in instantaneous OH radical concentration prediction. There is a shift in the location of maximum soot surface growth and oxidation rate with the non-gray radiation model. The radial variations of the soot volume fraction are shown in Fig. 13. The radial profiles are taken at the axial distance where the oxidation of soot is already started. It can be seen from the figure that the soot is significantly under predicted as we move away in the radial direction. We have observed earlier from figure 4 that the temperature at higher temperature is over predicted by ~200K, which favor's the oxidization and hence leads to the reduction of the net soot yield. Also, it is been discussed in the earlier section that including the non-gray radiation or even the turbulence radiation interaction, the drop in the global maximum temperature, which occurs ~r/D > 2 is minimal. Therefore, the soot volume fraction is not affected significantly by including the radiation heat transfer, which is evident from Fig 13. Also, as explained earlier the soot is oxidized quickly by the over predicted OH species. The mixture fraction based PDF gives relatively good prediction due to the over predicted average values. The predictions are improved with non-gray radiation model due to shift in location of maximum soot volume fraction.  Michael et al. [8] and Donde et al. [36] also observed an increase in the soot volume fraction when soot-turbulence-chemistry interactions were included using LES/PDF approach. The increase has been attributed to the sensitivity of the dissipation rate. The effect of soot-turbulence-chemistry on evolution of soot requires further considerations.

## 7. CONCLUSIONS

In this work soot formation in Delft III flame has been simulated using the Steady Laminar Flamelet model in conjunction with different soot models. First, the quality of the turbulence models is illustrated based on velocity and turbulent kinetic energy predictions. Apart from over prediction of mean mixture fraction, MSKE and RSM have predicted reasonable good match with experimental data. The velocity predictions showed good agreement with experimental data. The predictions of RSM model show improvement along the radial direction compared to MSKE making it the ideal choice for the soot predictions. The flamelet model over predicts the temperature. The soot predictions by Two Step and Moss-Brookes were found to be in reasonable match to the experimental data and much accurate than the One step model. One step model is not able to accurately predict the location of maximum soot formation due change in length of the flame caused by the turbulence model. Based on the particle density and soot volume fraction the soot evolution process can be classified into four stages. The numerical results obtained can be used to classify the process. For the Two step and Moss-Brookes for x/D < 50, the particle diameter and soot volume fraction are relatively small with respect to the number density which indicates nucleation of soot particles. For 50 < x/D < 95, the soot volume fraction increases for the Moss-Brookes indicating condensation of PAH onto existing particles. Finally, far downstream, the particles are oxidized by OH in the Moss-Brookes model.  For the Two step, the coagulation occurs further downstream (95 < x/D < 120). The effects of soot-radiation interaction on flame temperature and soot volume fraction are in good agreement with the theory. Turbulence-chemistry interactions are computed by Moss-Brookes model using the PDF of temperature or mixture fraction along with the soot scalars. Predictions of the PDF need to be accurate in order to capture infrequent sooting events. Due to inaccurate PDF predictions, the soot-turbulence interactions are not accurately captured. It is observed that the mixture fraction based PDF gives accurate results compared to the temperature based as it accounts for the variation in PAH formation due to turbulence.

**ACKNOWLEDGMENT**

The computation work has been carried out on the computers provided by IITK (www.iitk.ac.in/cc). Data analysis and article preparation has been carried out using the resources available at IITK. This support is deeply appreciated.

**NOMENCLATURE**

$a_s$      characteristic strain rate

$k_i$      absorption coefficient of gas

$k_{soot}$      absorption coefficient of soot

$a_{\epsilon,i}$      emissivity weighting factor

$b_{nuc}^*$      normalized radical nuclei concentration

$c_{p,i}$      $i^{th}$ species specific

$f$      mixture fraction

$p$      sum of the partial pressures of all absorbing gases

$R_{nuc}^*$      normalized rate of nuclei generation

$R_{soot}$      rate of soot formation

$T$      Temperature

$Y_i$      species mass fraction

$Y_{soot}$      soot mass fraction

$\kappa_i$      absorption coefficient of $i^{th}$ gray gas

$\rho$      Density

$\sigma_{soot}$      turbulent Prandtl number of soot transport

$\phi$      representative scalar

$\chi$      scalar dissipation rate

$\chi_{st}$      scalar dissipation rate at $f = f_{st}$

$erfc^{-1}$      inverse complementary error function

**Table Caption List**

Table 1          Experimental data of delft III burner [29]

**Figure Captions List**

Fig. 1          Top and cross view of axisymmetric non premixed jet burner

Fig. 2          Illustrative view of the mesh used [D: diameter of fuel jet]

Fig. 3        Centre line plots of axial velocity, temperature and mean mixture fraction using mske turbulence model: solid lines are coarse mesh, dashed lines are fine mesh and symbols are measurements.

Fig. 4        Radial profiles of axial velocity, turbulent kinetic energy, temperature and mean mixture fraction at three different locations from the fuel jet exit: lines are predictions and symbols are measurements

Fig. 5        Contours of mean temperature with (a) no radiation, (b) gray radiation and (c) non-gray radiation

Fig. 6        Radial profiles of mass fraction of oh, co, $h_2$ and $h_2o$ at three different locations from the fuel jet exit: lines are predictions and symbols are measurements

Fig. 7        Axial profile of soot volume fraction: solid lines are without radiation, dashed lines are with gray radiation, dashed dotted lines are with non-gray radiation and solid symbols are measurements

Fig. 8        Axial profile of soot volume fraction with different oxidation models: lines are predictions and symbols are measurements

Fig. 9        Contour of soot volume fraction showing (a) One step, (b) Two step and (c) Moss Brookes

Fig. 10       Contours of soot nucleation (left), soot surface growth (middle) and soot oxidation (right) using Moss Brookes model with (a) gray radiation and (b) non-gray radiation

Fig. 11       Axial profile of soot volume fraction with soot-turbulence interaction: solid lines are gray radiation, dashed lines are with non-gray radiation and solid symbols are measurements

Fig. 12       Axial profile of soot surface growth rate and soot oxidation rate using different presumed pdf for soot-turbulence interactions: solid lines are gray radiation and dashed lines are with non-gray radiation

Fig. 13    Radial profile of normalized soot volume fraction with soot-turbulence interactions: solid lines are gray radiation, dashed lines are with non-gray radiation and symbols are measurements

| | |
|---|---|
| $U_{fuel}$ | 21.9 m/sec |
| $T_{fuel}$ | 295K |
| $Re_{fuel}$ | 9700 |
| $U_{annulus}$ | 4.4 m/sec |
| $T_{annulus}$ | 295K |
| $Re_{annulus}$ | 8800 |
| $U_{coflow}$ | 0.4 m/sec |
| $T_{coflow}$ | 295K |

Table 1: Experimental data of delft III burner [29]

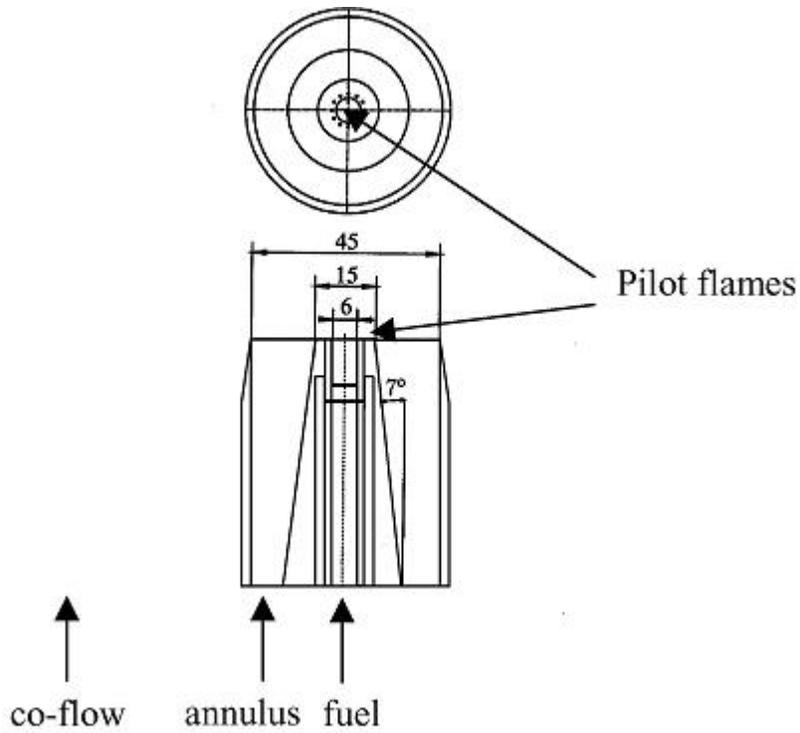

Figure 1: Top and cross view of axisymmetric non premixed jet burner [29]

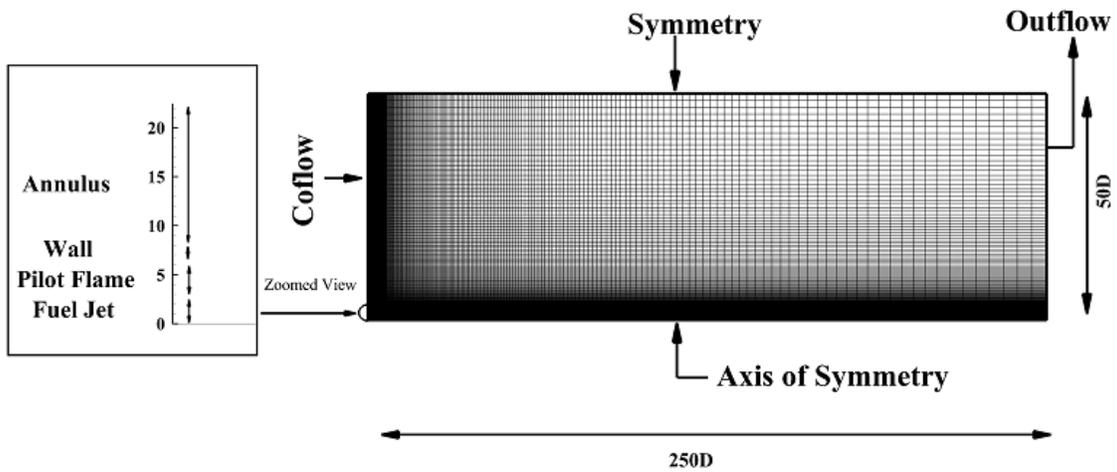

Figure 2: Illustrative view of the mesh used [D: diameter of fuel jet]

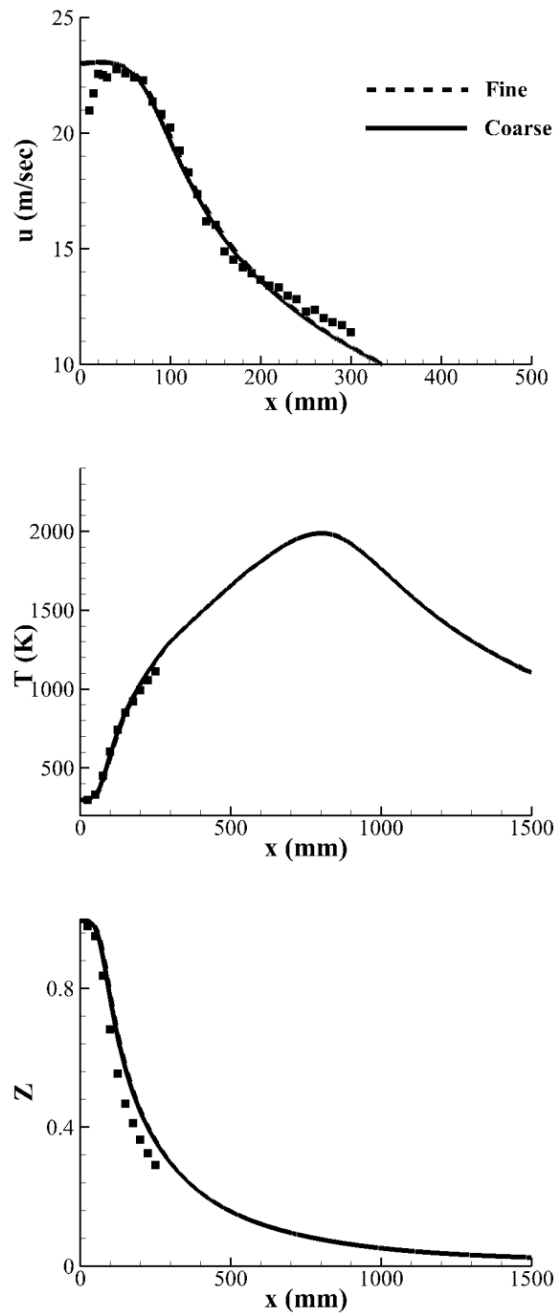

Figure 3: Centre line plots of axial velocity, temperature and mean mixture fraction using MSKE turbulence model: solid lines are coarse mesh, dashed lines are fine mesh and symbols are measurements [29]

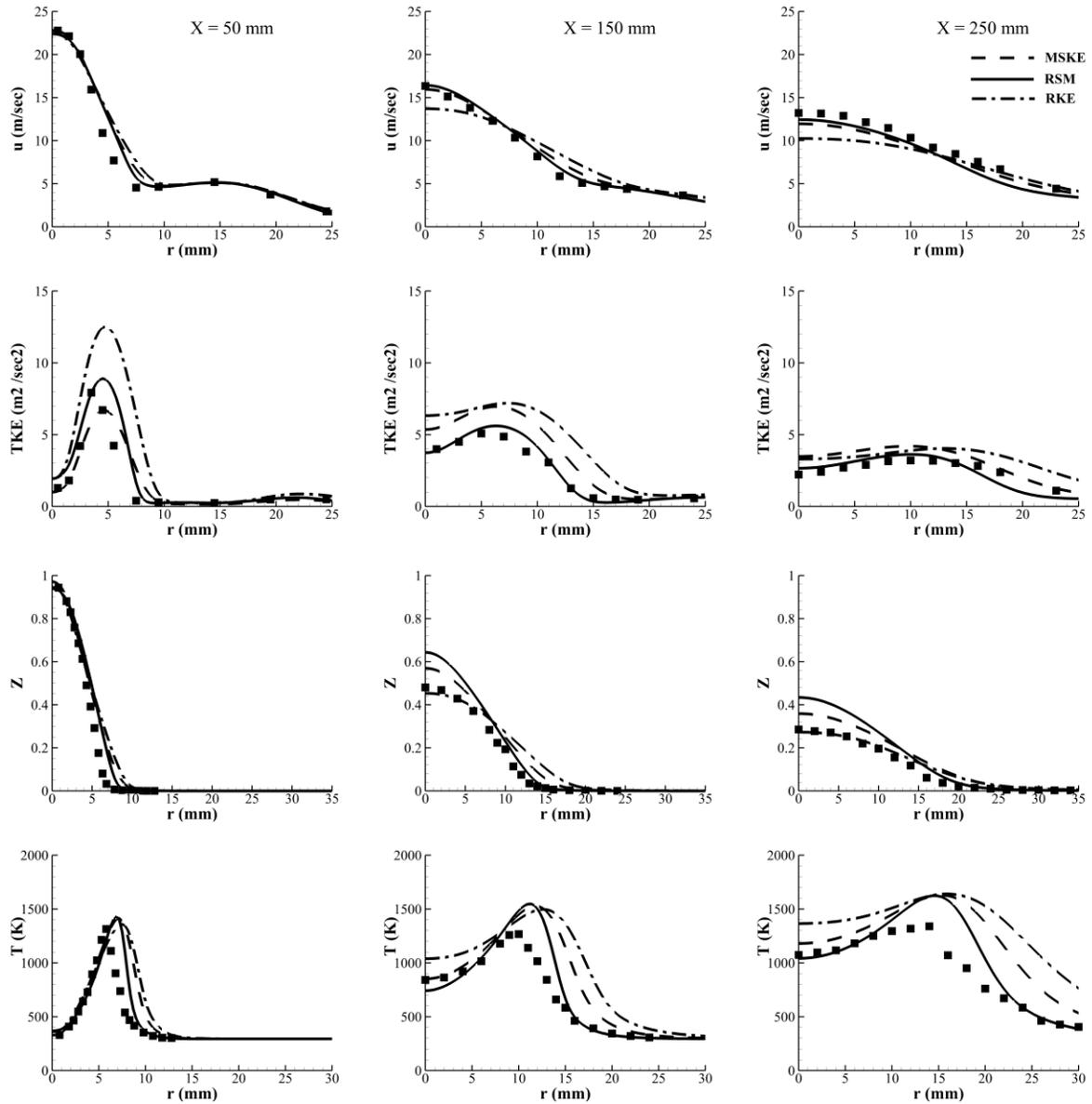

Figure 4: Radial profiles of axial velocity, turbulent kinetic energy, temperature and mean mixture fraction at three different locations from the fuel jet exit: lines are predictions and symbols are measurements [29]

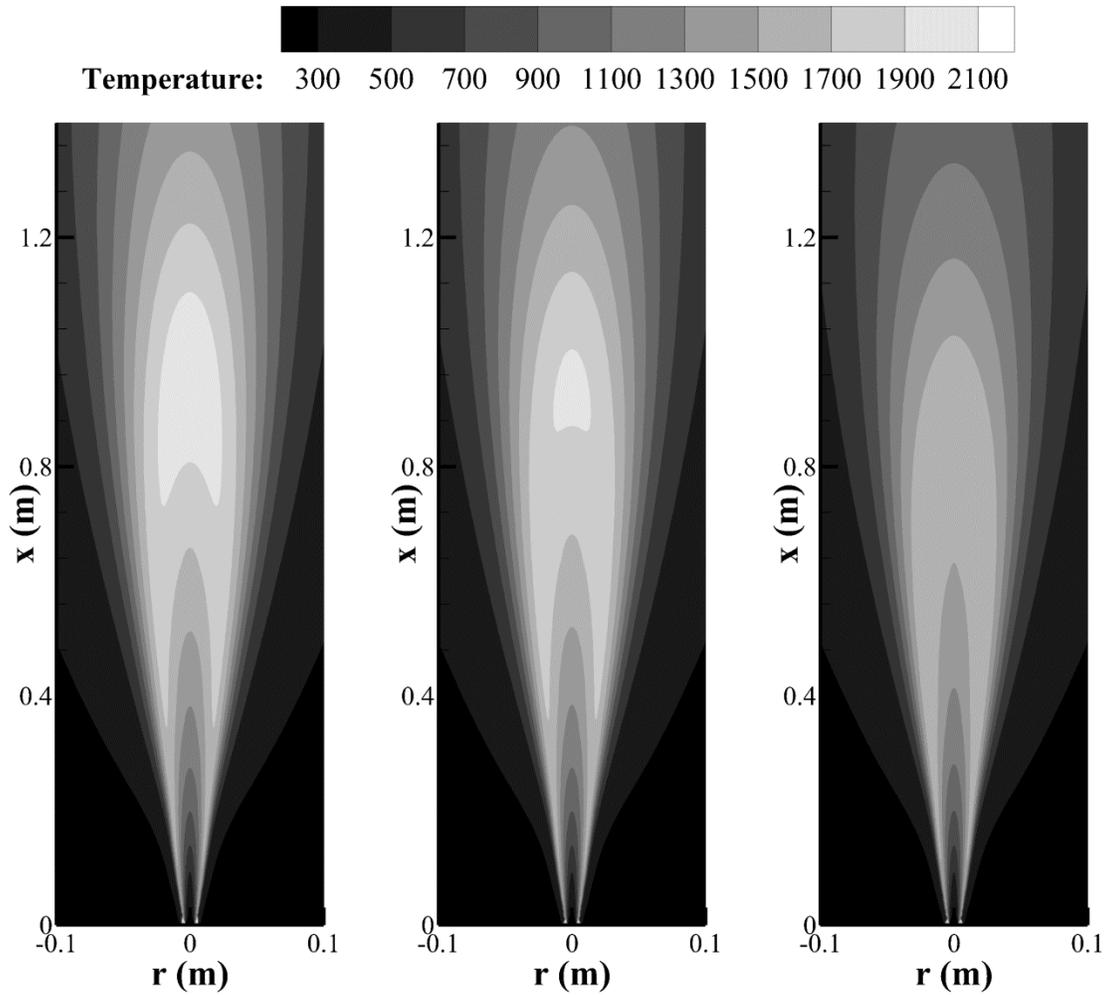

Figure 5: Contours of mean temperature with (a) no radiation, (b) gray radiation and (c) non-gray radiation

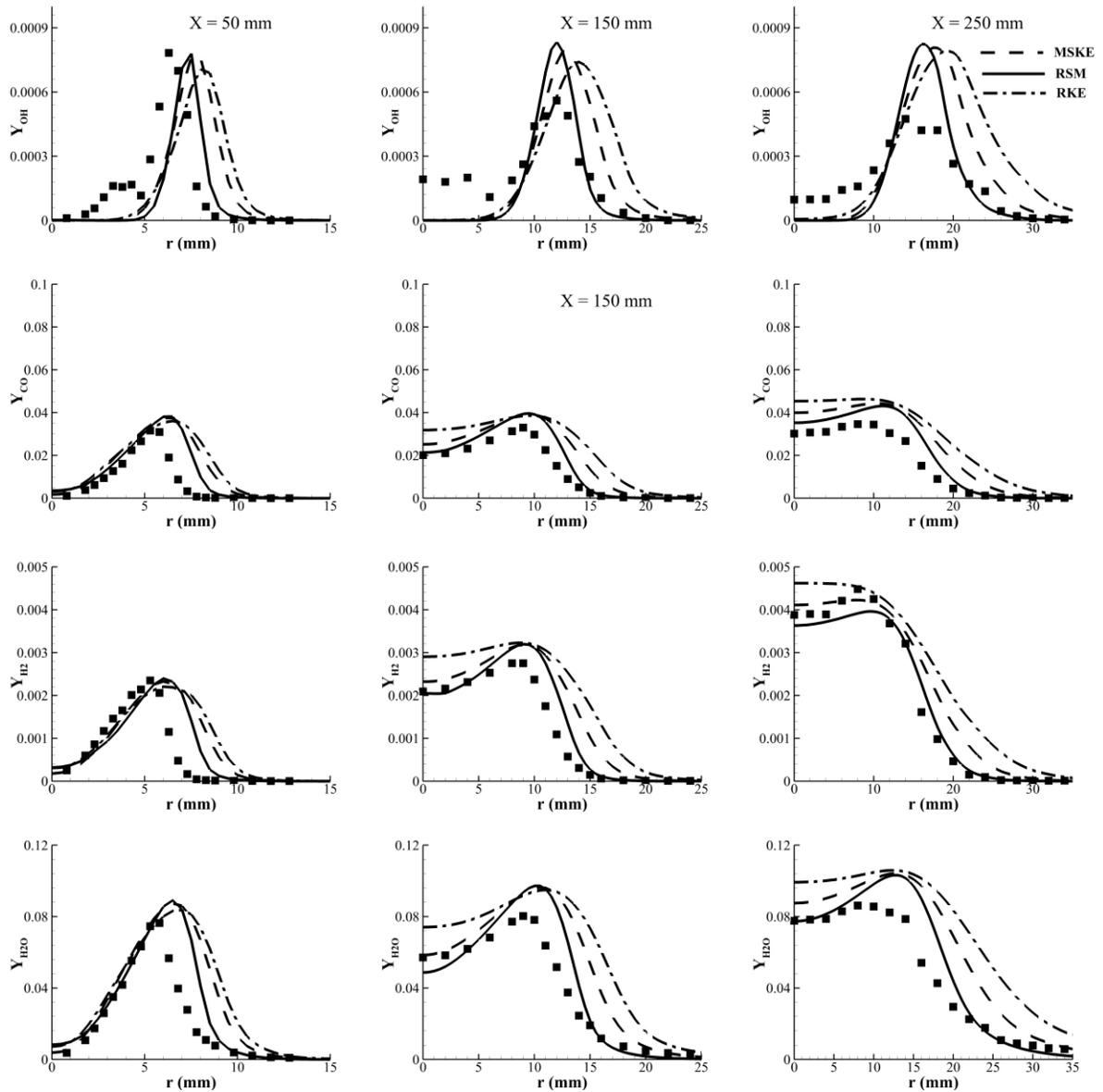

Figure 6: Radial profiles of mass fraction of oh, co, h$_2$ and h$_2$o at three different locations from the fuel jet exit: lines are predictions and symbols are measurements

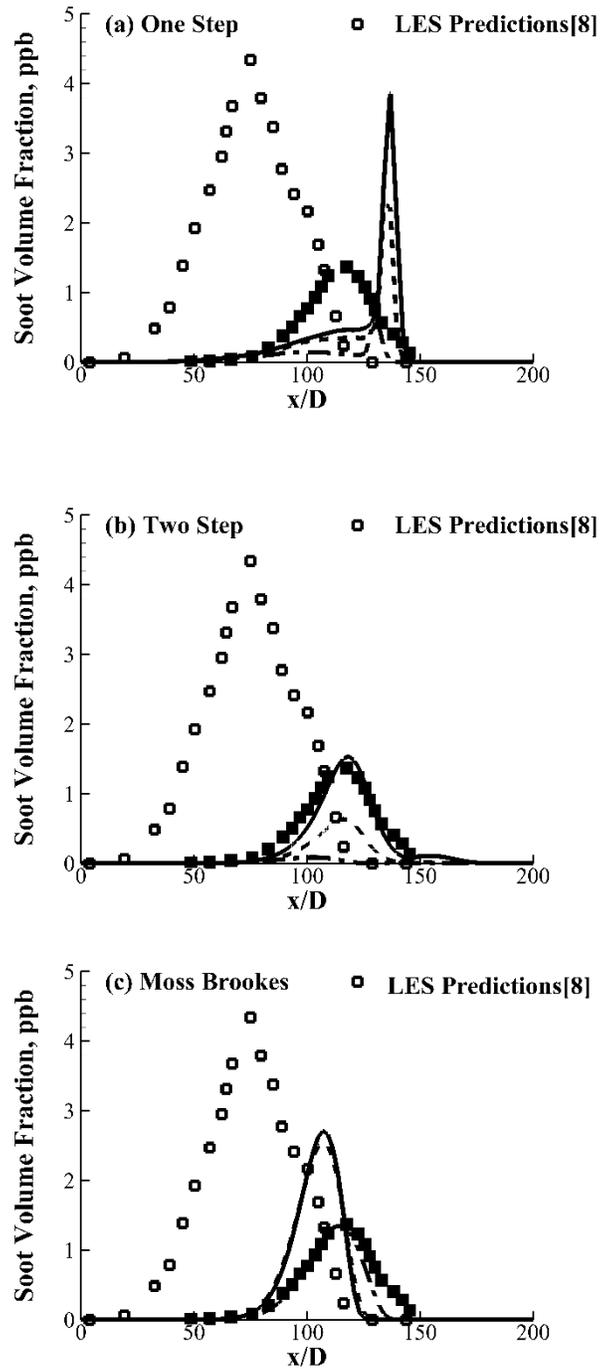

Figure 7: Axial profile of soot volume fraction: solid lines are without radiation, dashed lines are with gray radiation, dashed dotted lines are with non-gray radiation and solid symbols are measurements [9]

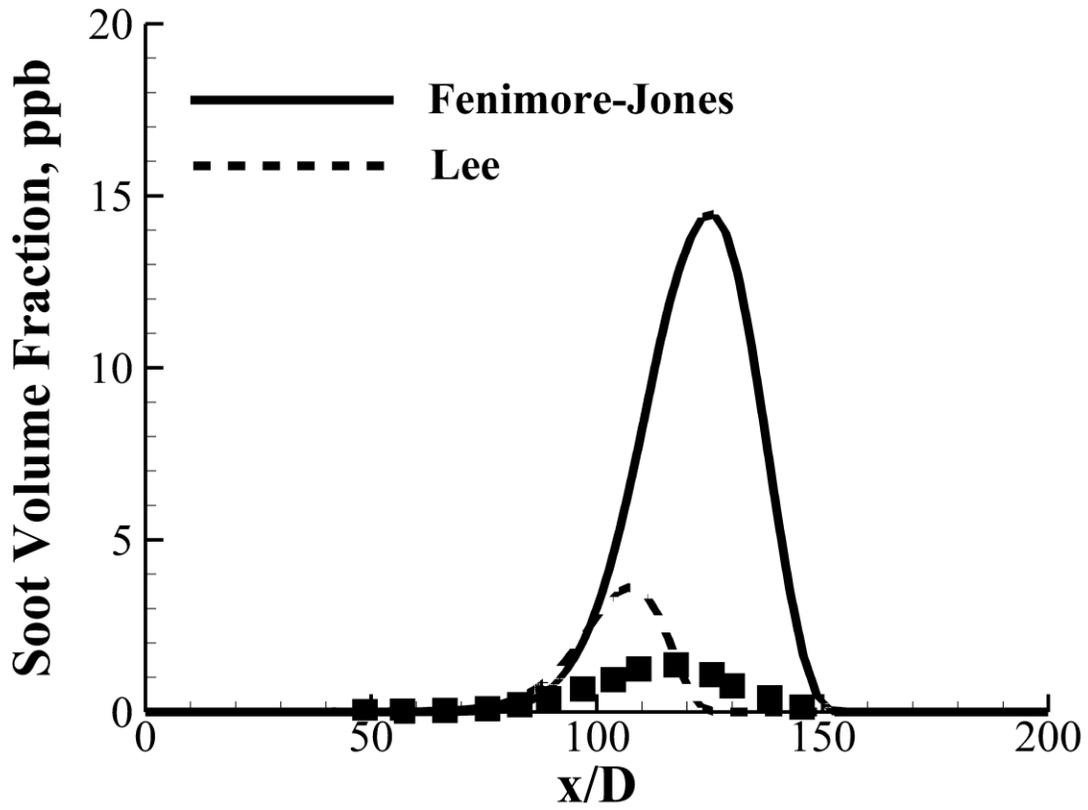

Figure 8: Axial profile of soot volume fraction with different oxidation models: lines are predictions and symbols are measurements [9]

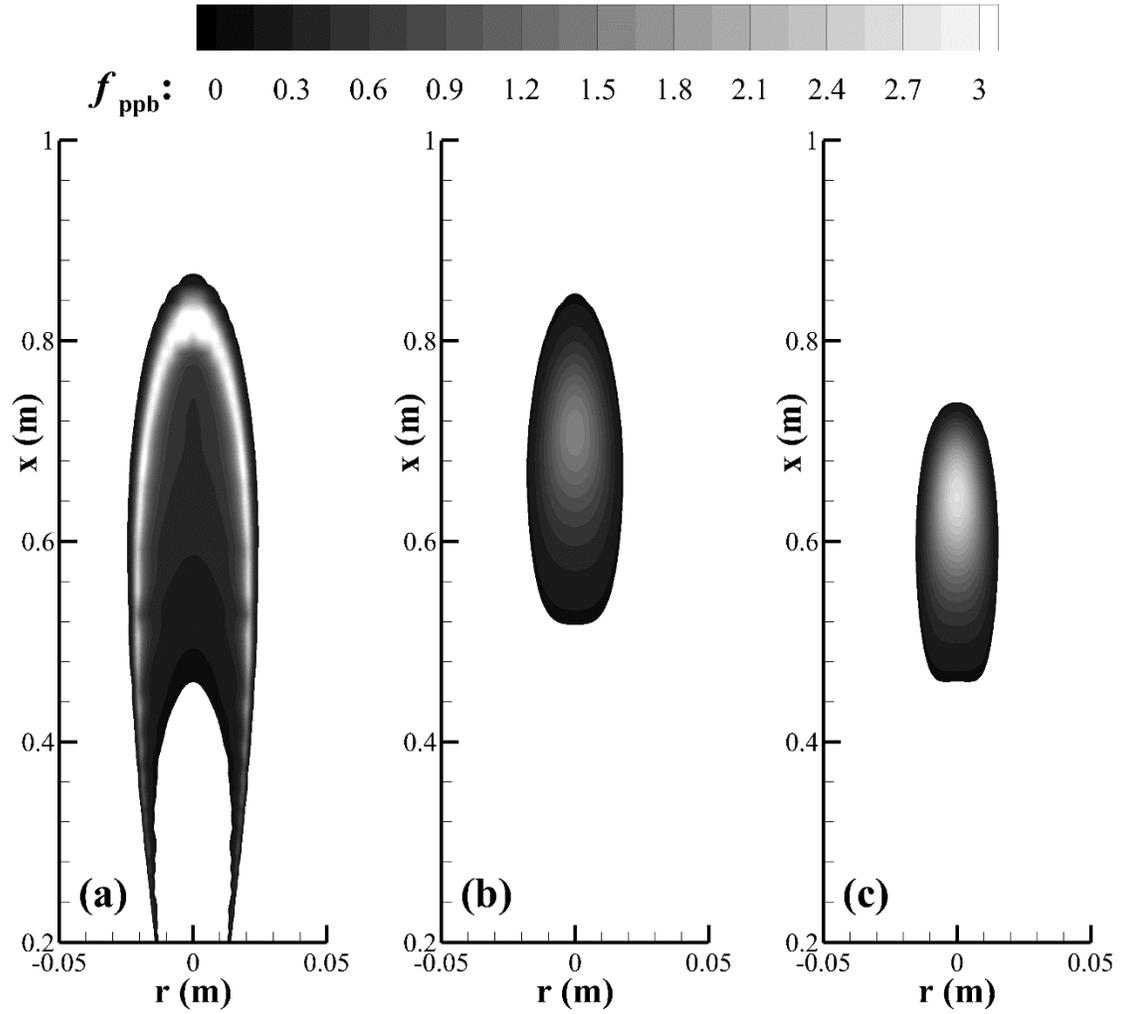

Figure 9: Contour of soot volume fraction showing (a) One step, (b) Two step and (c) Moss Brookes

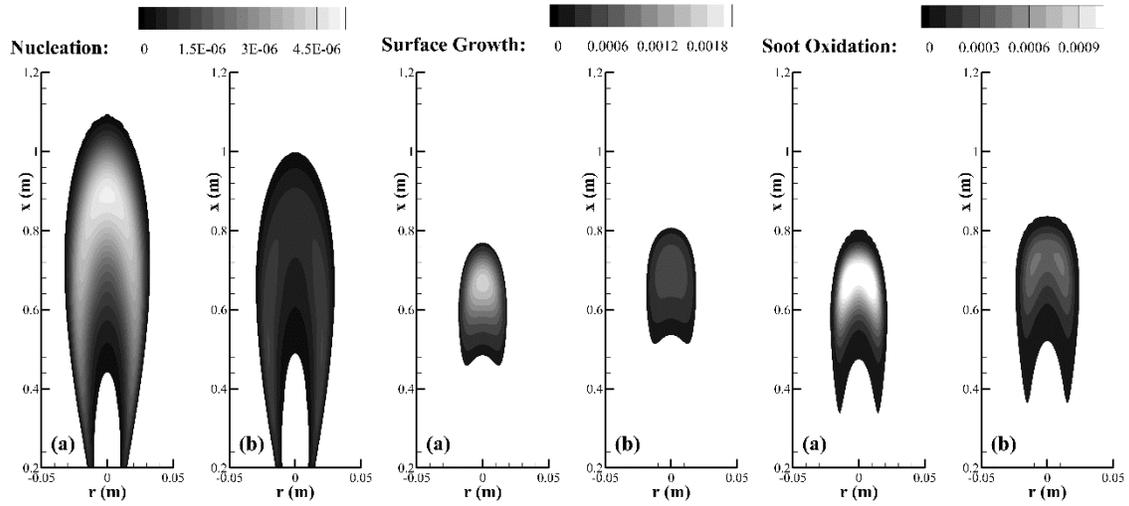

Figure 10: Contours of soot nucleation (left), soot surface growth (middle) and soot oxidation (right) using Moss Brookes model with (a) gray radiation and (b) non-gray radiation

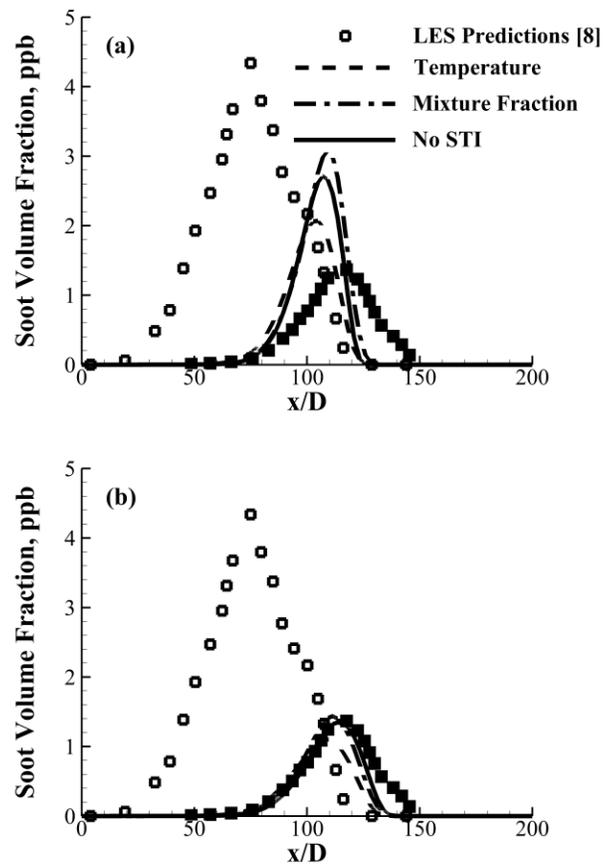

Figure 11: Axial profile of soot volume fraction with soot-turbulence interaction: (a) gray radiation, (b) non-gray radiation and solid symbols are measurements [9]

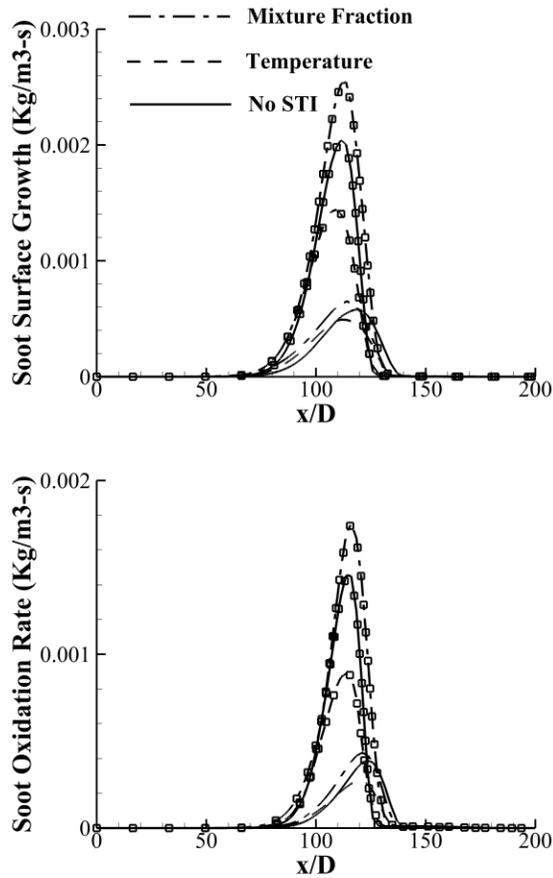

Figure 12: Axial profile of soot surface growth rate and soot oxidation rate using different presumed pdf for soot-turbulence interactions: lines with symbols are gray radiation and lines are with non-gray radiation

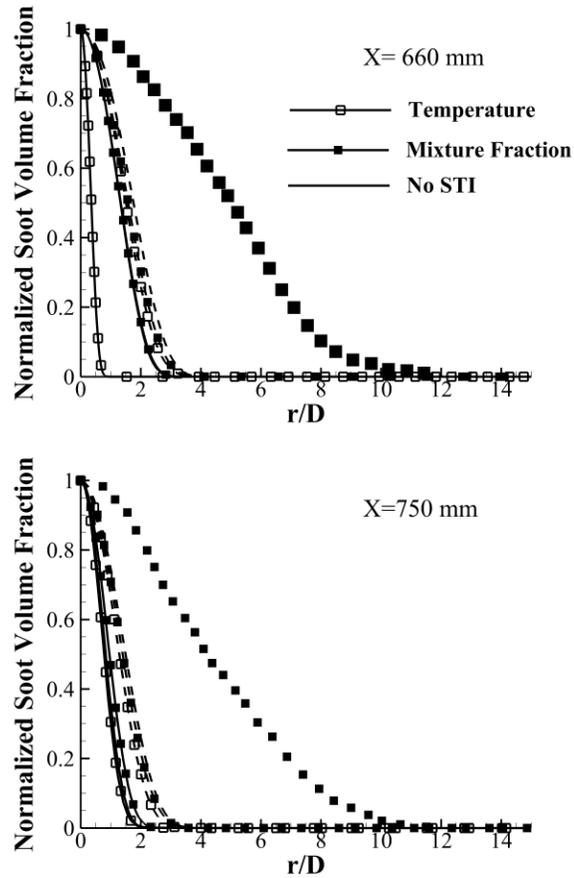

Figure 13: Radial profile of normalized soot volume fraction with soot-turbulence interactions: solid lines are gray radiation, dashed lines are with non-gray radiation and symbols are measurements [9]